\documentclass[10pt]{article}
\usepackage{latexsym,graphicx,amssymb,color,amsmath}
\usepackage{epsfig,lscape}

\begin{document}
\vspace*{-2cm}
\begin{flushright}
%\\
\end{flushright}

\vspace{0.3cm}

\begin{center}
{\Large {\bf Master equation approach to the assembly \\ \vspace{0.3cm}
of viral capsids }}\\ 
\vspace{1cm} {\large \bf T.\ Keef\,\footnote{\noindent E-mail: 
{\tt tk506@york.ac.uk}}, C.\
Micheletti\,\footnote{\noindent E-mail: {\tt michelet@sissa.it}} and
R.\ Twarock\,${}^{1,}$\footnote{\noindent E-mail: 
{\tt rt507@york.ac.uk}}}\\
\vspace{0.3cm} {${}^1$}\em Department of Mathematics \\ University of York\\
\vspace{0.3cm} {${}^3$} Department of Biology \\ University of York \\
York YO10 5DD, U.K.\\ \vspace{0.3cm} {${}^2$\em \it International School for Advanced Studies (S.I.S.S.A.)\\ 
 and INFM,  Via Beirut 2--4, \\ 34014 Trieste, Italy}\\ 
\end{center}

\begin{abstract}
The distribution of inequivalent geometries occurring during
self-assembly of the major capsid protein in thermodynamic equilibrium
is determined based on a master equation approach. These results are
implemented to characterize the assembly of SV40 virus and to obtain
information on the putative pathways controlling the progressive
build-up of the SV40 capsid. The experimental testability of the
predictions is assessed and an analysis of the geometries of the
assembly intermediates on the dominant pathways is used to identify
targets for antiviral drug design.
\end{abstract}

\section{Introduction}

Manipulating the assembly of viral capsids is one way of interfering
with the viral replication cycle and hence a possible avenue for
anti-viral drug design. Despite of its importance the theory of viral
capsid assembly is still in its infancy. A first model for the
self-assembly of a small plant virus was pioneered by Zlotnick
\cite{Zlotnick:1994}, exploring the assembly of a dodecagonal shape by
a cascade of single order reactions. It has since been extended to
more involved scenarios
\cite{Endres:2002,Zlotnick:1999,Zlotnick:2000}, including a study of
the energy landscape underlying assembly \cite{Endres:2005}, that is
similar to approaches in protein folding \cite{Brooks:1998} or the
energy landscape description of association reactions
\cite{Wales:1987,Wales:1996,Wolynes:1996}. These results have been
used to investigate the possibility of inhibiting assembly via an
anti-viral drug in the case of Herpes Virus
\cite{Zlotnick:2002}. Related approaches include molecular dynamics
studies of viral capsid assembly \cite{Rapaport:1999,Rapaport:2004},
and a molecular dynamics-like formalism that is implemented in
connection with a ``local rules" mechanism that regulates capsid
assembly \cite{Berger:1994,Schwartz:2000}.

A characteristic feature of these models is the fact that the bonding
structures of all building blocks are treated on an equal
footing. While this is justified for a large number of viruses, it is
an inappropriate simplification for important families of virus such
as the Papovaviridae, which are linked to cancer and are hence of
particular interest for the public health sector. For example, the
(pseudo-) T=7 capsids in this family are known to be composed of two
different types of pentameric building blocks which are distinguished
by their local bonding structure. This requires a mathematical
representation of these building blocks that takes the differences in
the local bonding environments into account. The tiling approach for
the description of viral capsids \cite{Twarock:2004a,Twarock:2004b}
provides an appropriate mathematical framework for this. It encodes
the locations of proteins and inter-subunit bonds in terms of tilings,
i.e. tessellations that represent the surface structure of the
capsids. Since the vertex atlas of these tilings, i.e. the collection
of all distinct local configurations around vertices in the tiling,
encodes all different types of capsomeres and their bonding
structures, it provides appropriate building blocks for the
construction of assembly models. An assembly model for (pseudo-) T=7
capsids in the family of Papovaviridae has been introduced along these
lines in \cite{KTT}. In this reference, a tree structure -- the
assembly tree -- has been determined that encodes all energetically
preferred pathways of assembly, and it specifies how the tree
structure changes in dependence on the association
constants. Moreover, the concentrations of the statistically dominant
assembly intermediates (i.e. inequivalent shapes at various stages of
capsid construction that are located on all pathways) have been
computed.

For applications to anti-viral drug design, it is important to control
not only the statistically dominant assembly intermediates, but also
the concentrations of all other assembly intermediates, and, based on
this, to determine the most probable assembly pathway(s). This issue
is addressed in this paper, where we adopt a master equation approach
for the computation of the concentrations of all assembly
intermediates in the assembly tree and use this information to
determine the putative pathways controlling the progressive build-up
of the capsid. In particular, in section \ref{two} we introduce the
master equation approach as a tool for the computation of the
concentrations of the assembly intermediates for arbitrary viruses
from a thermodynamical point of view. In section \ref{three} we apply
this formalism to SV40 virus and determine the equilibrium
concentrations of the various assembly intermediates. We discuss how
this information can be used to determine the dominant pathway of
assembly, and show that the dominant pathways have intermediates with
a characteristic structure that may potentially be exploited in the
framework of anti-viral drug design. In the final section we summarize
our results and assess the implications for other families of viruses.

\section{The master equation approach}\label{two}

The formalism presented in this section allows to compute the
probability distribution of the inequivalent configurations (also
called species or assembly intermediates) that appear during
self-assembly of the major capsid protein of a virus in thermodynamic
equilibrium. Assume that the different species are indexed from 1 to
$N$, where species 1 corresponds to the fundamental building block of
the capsid, $N$ to the final capsid, and every other assembly
intermediate, $i$, is formed by $n_i$ copies of building block 1. As
customary, we consider capsid assembly as a sequence of low order
reactions, and hence assume that the formation of the capsid occurs
from the attachment or detachment of single fundamental units to the
partially-formed capsid. From a phenomenological point of
view, the equilibrium thermodynamics of the process is described
through second-order association constants. Indicating with $[a]$ and
$[b]$ the concentrations of two species whose number of constitutive
building blocks is, respectively, $n_a$ and $n_b = n_a+1$, we have
that their association constant is given by 

\begin{equation}
K_{b,a} = { [b]_{eq} \over [a]_{eq} [1]_{eq}}\,.
\label{eqn:secord}
\end{equation}
 
\noindent where $[1]$ denotes the concentration of the fundamental
building block and the subscripts are used to stress that the
concentrations pertaint to the equilibrium (stationary) state. This
phenomenological relationship can be related to the fundamental
entropic and energetic aspects of the association process through the
following factorization (as in \cite{Zlotnick:1994}, \cite{KTT}):
\begin{equation}
K_{b,a} = \frac{1}{c_0} S_1 \, S_{ba}\, e^{- \Delta G(b,a) \over R
T}\,,
\label{eqn:k}
\end{equation}
where $S_1$ denotes the geometric degeneracy of the fundamental
``incoming'' subunit, $S_{ba}:= O(b)/O(a)$ corresponds to the ratio of
the orders of the discrete rotational symmetry groups of the two
species $O(b)$ and $O(a)$, $\Delta G(b,a)$ is the free energy
difference associated to the bonds formed by the incoming building
block, and $R$ and $T$ denote the gas constant ($R=1.987 cal K^{-1}
mol^{-1}$) and, respectively, temperature (chosen as room temperature
$T=298 K$). The quantity $c_0$, having the dimension of a
concentration, can be put in unique correspondence with the total
concentration of elementary blocks present in the system, as will be
shown below.

\noindent The hierarchy of association constants of the various pairs
of intermediate species differing by one fundamental building block can be
used in recursive schemes for obtaining the equilibrium probabilities
of any species. In fact, by combining equations (\ref{eqn:secord}) and
(\ref{eqn:k}) and introducing the adimensional quantity
${\widetilde{[1]}_{eq}} := \frac{[1]_{eq}}{c_0}$ we obtain the
fundamental relationship:

 \begin{equation}
{ [b]_{eq} \over [a]_{eq}} = S_{1} \, S_{ba}\,
  {\widetilde{[1]}_{eq}}\, e^{- \Delta G(b,a) \over R T}
\label{eqn:ratio}
\end{equation}

\noindent which, used recursively, yields the formal expression of the
equilibrium concentration of a generic species, $[i]$, ($i \not= 1$), 

\begin{equation}\label{ieq}
[i]_{eq} = \displaystyle S_1^{n_i-1} \frac{O(1)}{O(i)} e^{-
\frac{\Delta G(i,1)}{R T}}\,  [1]_{eq} \,{\widetilde{[1]}_{eq}}^{n_i-1},
\end{equation}
\noindent where $n_i$ is the number of fundamental building blocks in species
$i$ and $O(1)$ and $O(i)$ refer to the orders of discrete rotational
symmetry of subunit $1$ and species $i$, respectively. Notice that
equation (\ref{ieq}) depends implicitly on $c_0$ through
$\widetilde{[1]}_{eq}$. This gives the possibility of relating $c_0$
to the total concentration of fundamental building blocks, $[c^*]$, through the
relationship,

\begin{equation}\label{eqn:conslaw}
[c^*]= \sum_{i=1}^N \, n_i\, [i]\ .
\end{equation}

\noindent Notice that this relationship expresses the law of
conservation of the total number of fundamental building blocks present in the
system (be they ``free'' or assembled in intermediate species) and
therefore is valid not only in equilibrium. The association constants
of equation (\ref{eqn:secord}) can be used beyond the equilibrium
framework since they constitute the starting point for formulating
phenomenological kinetic equations apt to capture the time evolution
of the system given the initial concentration of the various species.
Within a vanishingly small time interval the concentration of a given
species $[i]$ (we assume $i \not= 1$) can change only due to the gain
or loss of one fundamental building block:

\begin{equation}
{d [i]\over d t} = \sum_m\, [m]\, W_{m,i}+ \sum_l\, [l] [1]\, W_{l,i} - \sum_m\, [i] [1]\,
W_{i,m} - \sum_l\, [i]\, W_{i,l}
\label{eqn:conckinetics}
\end{equation}

\noindent where we have omitted the explicit time dependence of the
species concentrations. The indices $l$ and $m$ in the sums refer to
the species formed by one less, respectively one more, fundamental
building block than species $i$. Finally, $W_{i,j}$ denotes the {\em
time-independent} rate at which transitions are made from
configuration $i$ to configuration $j$. The dynamics of the system is
thus fully described by the set of coupled equations
(\ref{eqn:conckinetics}) for each $i \not= 1$, supplemented with the
conservation law in equation (\ref{eqn:conslaw}). The $W$'s must be
appropriately related to the association constants, $K$, to ensure
that the correct equilibrium conditions (\ref{eqn:secord}) are
recovered at large times when the stationary regime is reached
(i.e. when $d [i]/ d t=0$ for all species $i$).

Among all possible initial conditions for the above mentioned kinetic
evolution a particularly appealing and interesting one is represented
by the case where the only species being present is the one associated
with the fundamental building blocks. At this initial time ($t=0$) the state of
the system would then be described as $[1]_{t=0}=[c^*]$ and
$[j]_{t=0}=0$ for all species $j>1$. The lack of precise experimental
characterization of the equilibrium concentrations of the various
species in biologically-relevant conditions has lead previous
theoretical studies to focus on particularly simple equilibrium
situations, namely the one in which the only dominant species are that
of the fully-formed capsid and that of the fundamental building
blocks; both species are considered as equiprobable so that
$[N]_{eq}=[1]_{eq}$ while, for all other species $j$,
$[j]_{eq} \approx 0$.

Under these assumptions, the concentration of the fundamental species
[1], is therefore expected to take on a rather limited range of
values. We build on this observation to simplify the description of
the assembly process of equation (\ref{eqn:conckinetics}) through a set of
effective first order reactions. The key ingredient in our analysis is
to modify the right hand side of equation (\ref{eqn:conckinetics}) so to
neglect the time-dependence of the concentration $[1]$ and absorb it
in new effective {\em time-independent} transition rates, $W$.  It is
convenient to recast the kinetics obtained by this simplification of
equation (\ref{eqn:conckinetics}) not in terms of the concentration of the
$i$th species but of the equivalent probability of occurrence, $P_i(t)
= [i] / \sum_j [j]$. The discrete time evolution of $P_i$ is therefore
governed by the following master equation

\begin{equation}
P_i(t+ \Delta t) = P_i(t)  + \Delta t \left(  \sum_j P_j(t) \, W_{j,i} -
\sum_j P_i(t) \, W_{i,j} \right)
\label{eqn:master}
\end{equation}
\noindent where $\Delta t$ is the time step of the discretized
evolution (assumed to be sufficiently small to justify the
linearization of the continuous time evolution). As before, the only
non-zero entries in the transition matrix $W$ are those connecting
species which differ by the addition/removal of one fundamental
building block. The matrix $W=(W_{i,j})$ has to satisfy a number of
properties (see \cite{Itzykson}):

\begin{itemize}
\item $\Delta t\, \sum_j W_{i,j}=1$\ \ \  (normalization condition)\ ,
\item there exists a finite integer $l$ such that $[W^l]_{i,j} > 0 \ \forall i,j$
  \ \  (ergodic condition)\,.
\end{itemize}

It is easy to check that the first condition ensures that $\sum
P_j(t)$ is constant at all times while the second one requires that
any two configurations must be connected by a finite number of
transitions.  The above conditions are sufficient to ensure the onset
of equilibrium at $t \to \infty$, irrespective of the initial
condition of the system.  From the stationarity of the equilibrium
distribution we obtain, from equation (\ref{eqn:master}) the generalised
balance condition:
\begin{equation}
 \sum_j ( P_j^{eq} \, W_{j,i} - P_i^{eq} \, W_{i,j} ) =0 \ .
\label{eqn:gb}
\end{equation}

\noindent The constraints entailed by the balance equation are not
sufficient to identify the matrix $W$ uniquely. We solve this
ambiguity by adopting the commonly-employed Metropolis criterion
within a detailed balance scheme \cite{Itzykson}. The detailed balance
condition requires that each term in the sum of eqn. (\ref{eqn:gb})
is zero. The Metropolis criterion further specifies the precise form
of the $W$ matrix elements. Accordingly, for two different species $i$
and $j$, which differ by the addition/removal of one fundamental
building block (otherwise $W_{ij}=0$) one has

\begin{equation}
W_{i,j} = \left\{
\begin{array}{l l}
 1 & \mbox{if  $P^{eq}_i < P^{eq}_j$}\\
 P_j^{eq}/P_i^{eq} & \mbox{otherwise}
\end{array}
\right.\,.
\label{eqn:metropolis}
\end{equation}

\noindent The diagonal elements are instead obtained from the
normalization condition:

\begin{equation}
W_{ii} = {1 \over \Delta t} - \sum_{j \not= i}\, W_{i,j}\,.
\end{equation}

\noindent It is easy to check that with this choice of $W$ the
equilibrium distribution is stationary under the evolution of
equation (\ref{eqn:master}). In the particular context of capsid
assembly, the ratio of probabilities in
equation (\ref{eqn:metropolis}) is straightforwardly obtained from
equation (\ref{eqn:ratio}) given the proportionality of $[i]$ and $P_i$.

Some caveats must be borne in mind when interpreting the outcome of
the master equation as a kinetic process. While the equilibrium
distribution is insensitive to the choice of the $W$'s, as long as
they satisfy equation (\ref{eqn:gb}), the kinetics is strongly affected by
the form of the $W$ matrix. Our choice follows the common practice of
adopting the Metropolis criterion, but remains only one of the
equivalent possibilities in terms of correct asymptotic behaviour.
Also, we stress again that recasting equation (\ref{eqn:conckinetics})
into the master equation of equation (\ref{eqn:master}) was possible upon
neglecting the time dependence of [1] in (\ref{eqn:conckinetics}).

\section{Application of the formalism to SV40 virus}\label{three}

\begin{figure}[ht]
\begin{center}
\includegraphics[width=3.8cm,keepaspectratio]{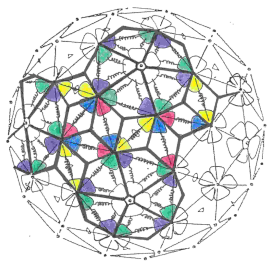}
\end{center}
\caption{The tiling representing the viral capsid of SV40.}
\label{Tiling}
\end{figure}

\begin{figure}[ht]
\begin{center}
\includegraphics[width=3.8cm,keepaspectratio]{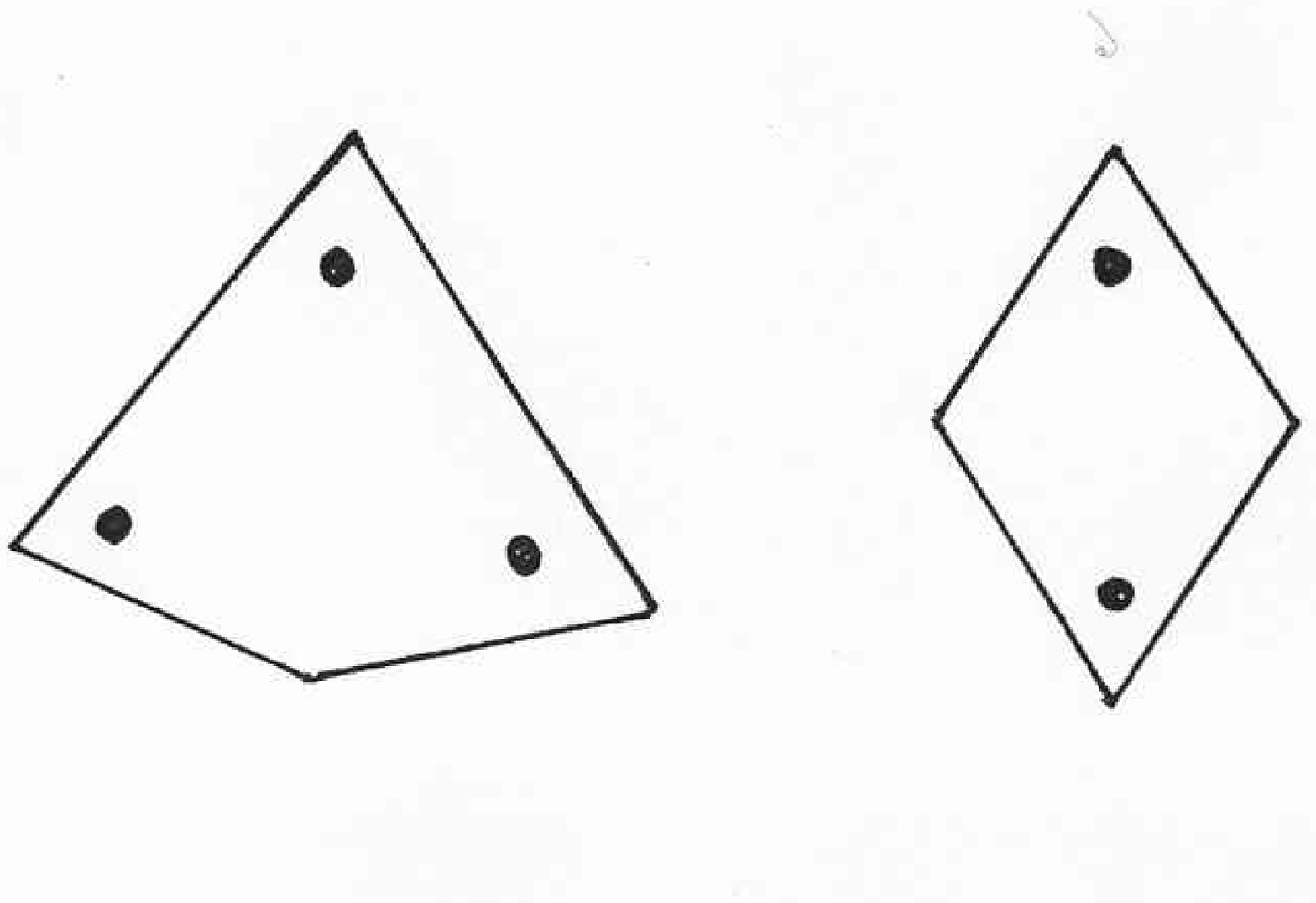}
\end{center}
\caption{Tiles corresponding to the tiling in Fig. \ref{Tiling}.}
\label{Tiles}
\end{figure}

In this section we apply the master equation formalism to the assembly
of SV40 virus. The capsid of SV40 is composed of 72 pentameric
building blocks that adopt two different types of local configurations
(with respect to their local bonding structure) in the capsid. These
can be modelled via the tiling approach shown in Fig. \ref{Tiling}
adapted from \cite{Twarock:2004a}.
%----------------------------------------- Figure --------------------
%--------------------------------------------------------------------
In particular, dimer- and trimer interactions, that is interactions
between two, respectively three, protein subunits, are represented
geometrically as rhombs, respectively kites. These rhombs and kites (see also Fig. \ref{Tiles}) 
tessellate the surface of the capsid, and encode the locations of the
protein subunits and the inter-subunit bonds. 
%----------------------------------------- Figure --------------------
%--------------------------------------------------------------------
In particular, tiles have to be interpreted as follows: dots on the
tiles correspond to angles of equal magnitude and mark the locations
of the protein subunits. The locations of the inter-subunit bonds
(C-terminal arm extensions in the case of SV40) correspond to the
straight lines connecting these dots.

\begin{figure}[ht]
\begin{center}
\includegraphics[width=5cm,keepaspectratio]{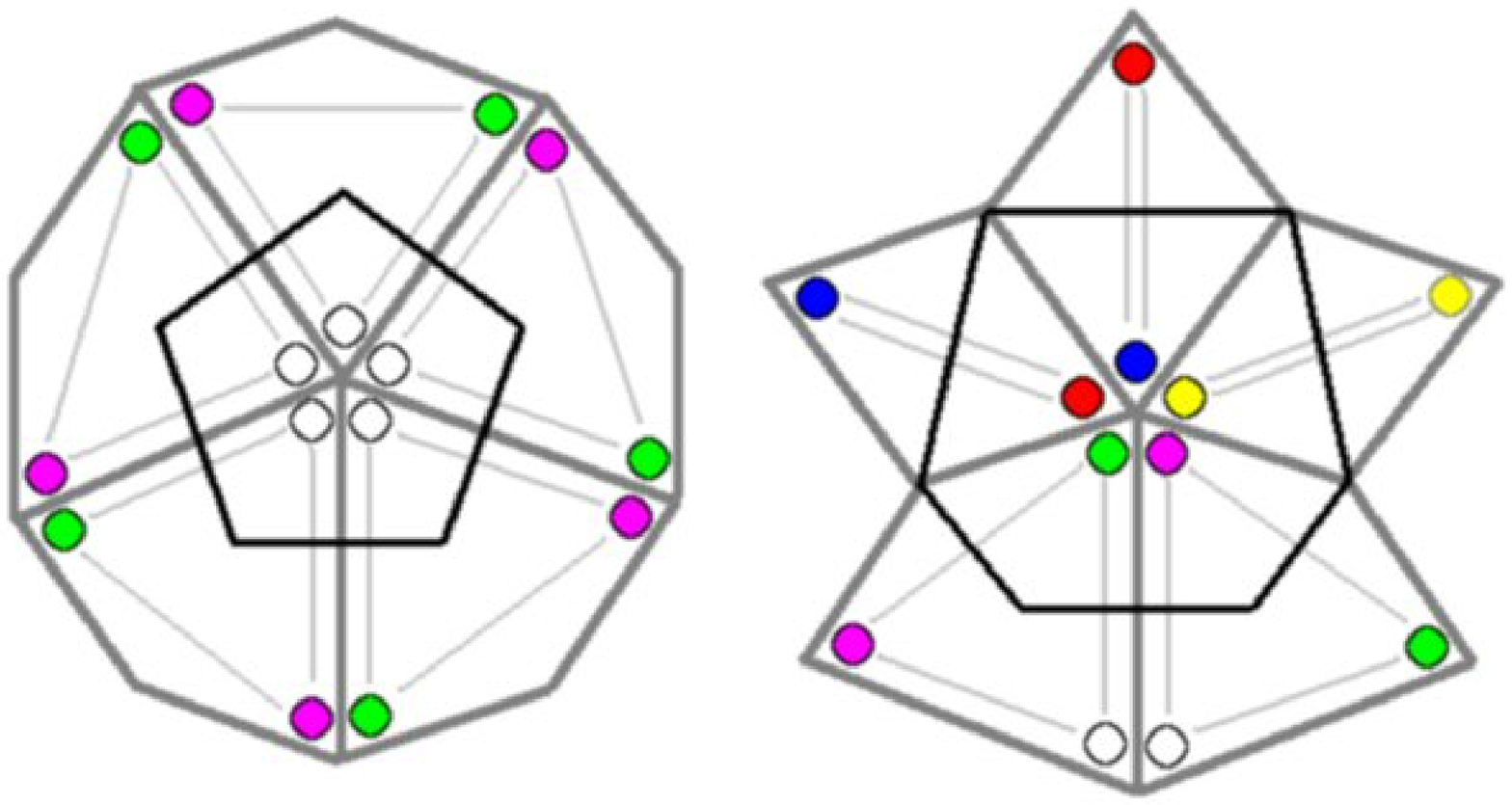}
\includegraphics[width=5cm,keepaspectratio]{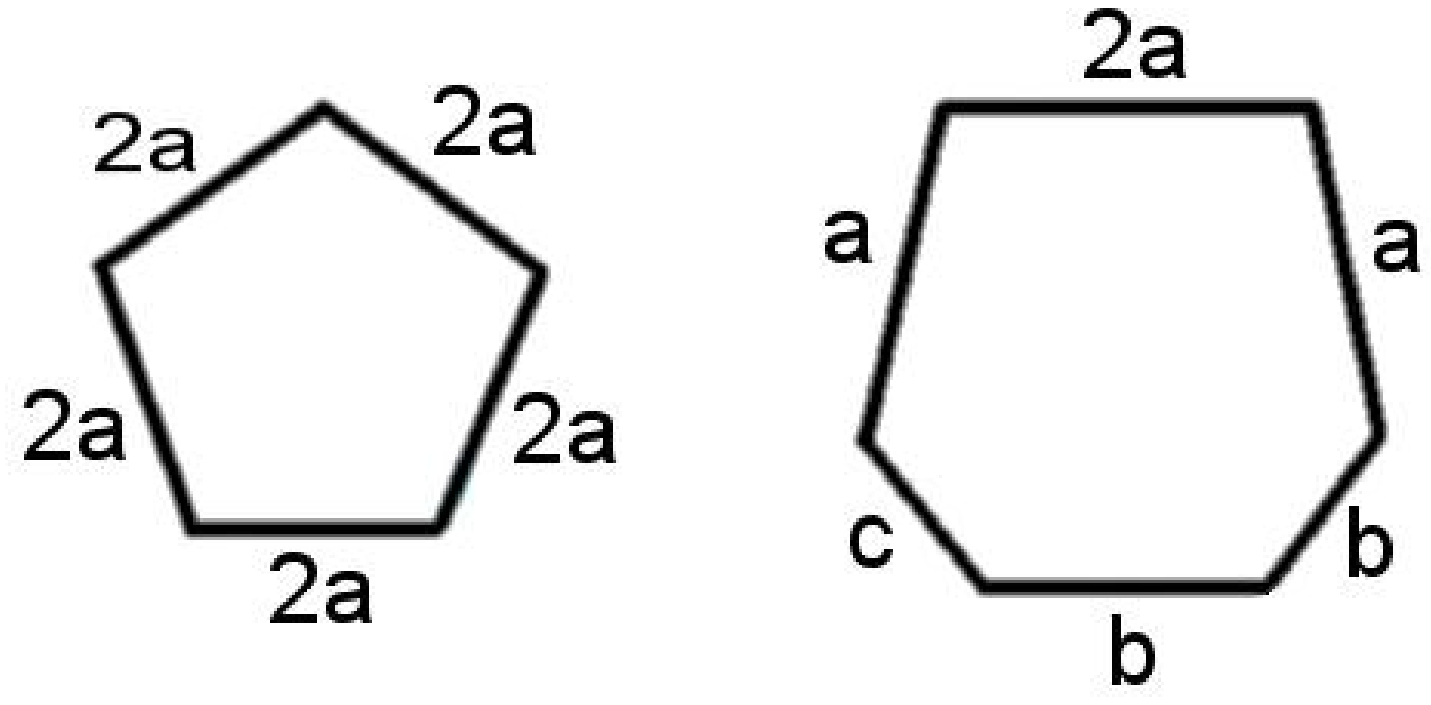}
\end{center}
\caption{The building blocks for SV40 capsid assembly.}
\label{BuildingBlocks}
\end{figure}

SV40 has an icosahedrally symmetric capsid, and therefore also the tiling has this
symmetry. From the tiling, one can see that the twelve pentamers
located at the 5-fold vertices of the capsid are surrounded by
trimer-interactions, and the 60 other pentamers by a combination of
dimer- and trimer-interactions. There are hence two different types of
local environments, which are shown in Fig. \ref{BuildingBlocks}.
%----------------------------------------- Figure --------------------
%--------------------------------------------------------------------

Instead of using the tiles themselves, it is more convenient -- and
mathematically equivalent -- to work with the pentagons and hexagons
shown superimposed on the tiles (Fig. \ref{BuildingBlocks}, left),
where the edges of the pentagons and hexagons are labelled according
to the association energies related to the tiles they bisect (same
figure, right). In particular, for SV40 there are 3 different types of
bonds, that correspond to the association energies $a$ for a single
C-terminal arm in a kite tile, association energy $b$ for a
quasi-dimer bond ("yellow-yellow" rhombs in Fig. \ref{Tiling}, named
after their location on a local 2-fold symmetry axis), and $c$ for a
strict dimer bond ("blue-red" rhombs in the figure, named after their
location on a 2-fold symmetry axis). They have to be taken into
account when computing the free energies $\Delta G$ in
equation~\ref{eqn:ratio}.

In \cite{KTT} the assembly of SV40 virus is considered based on these
pentagonal and hexagonal building blocks.  Assembly is considered as a
cascade of low order reactions by association of a single building
block at a time. The complete characterization of the assembly
thermodynamics would require the full classification of all possible
species of correctly-connected pentamers and hexamers, that is all
combinatorially possible combinations of these building blocks, even
those comprising more than the 72 blocks blocks necessary to form the
full capsid (and may correspond to malformed capcids or other types of
closed or open structures). Obviously, this exhaustive enumeration
cannot be accomplished. Consequently it is necessary to reduce the
number of building blocks to a manageable size by discarding
configurations that are supposed to play an unimportant role in the
assembly process. To illustrate the reduction procedure employed in
this study it is convenient to regard the various species as the nodes
of a graph. The links in the graph connect species which differ by the
attachment/removal of a pentamer of hexamer. The assembly tree
containing the species to which we restrict our attention is
constructed as follows. Without loss of generality the first node is
constituted by the fundamental pentamer. We then consider all the
geometrically inequivalent species obtained via the addition of an
extra building unit. Of these species we retain only the one (or ones
in case of degeneracy) having the lowest free energy. Each of the
retained species become new nodes of the graph (and are linked to the
parent node). In correspondence of each of these offspring nodes we
carry out the search of minimum free-energy descendants, as
before. The process ultimately ends when the node corresponding to the
full capsid is reached.

Within this limited set of species, the possible assembly pathways are
represented as walks on the tree connecting linked nodes. We stress
again that, due to the selection criterion based on the free energy
minimization, the assembly tree that we obtain represents only a
subset of the combinatorially possible nodes. The resulting ``minimal
tree'' is uniquely encoded by the energy parameters, $a,\ b$ and
$c$. Since one of these three quantities can be taken as the unit of
energy, we have that the choice of the assembly tree depends on two
adimensional parameters: $a/c$ and $b/c$. In \cite{KTT}, the phase
diagram of the system corresponding to these parameters is
depicted. It is shown that one can identify convex regions in this
two-dimensional parameter space where parameters can be varied without
affecting the assembly tree (but, obviously, the probability of
occurrence of the various species will change from point to point in
the same region).

While no direct measurement of the association energies $a$, $b$ and
$c$ is available, an estimate for the assembly free energies is
provided by the VIPER database. We discuss the assembly tree
associated with the point $x\equiv b/c \approx 0.92$ and $y\equiv a/c
\approx 0.47$, in the phase space, as it corresponds to the {\em
ratios} of the association energies listed on the VIPER webpages for
SV40.

\begin{figure}[h]
\begin{center}
\includegraphics[width=5cm,keepaspectratio]{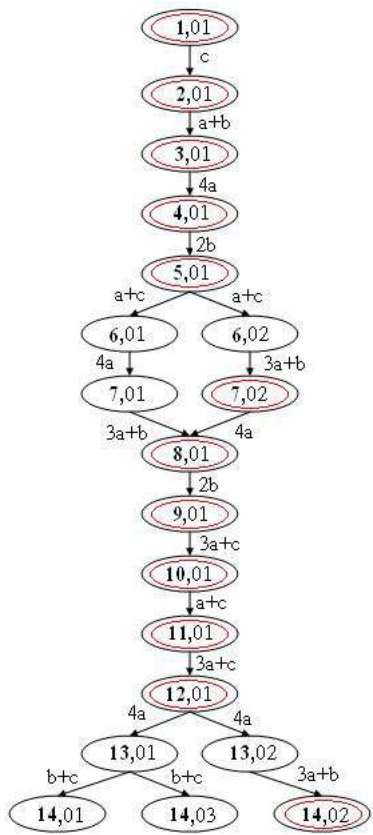}
\end{center}
\caption{The start of the assembly tree for SV40.}
\label{figSV40}
\end{figure}

In Fig. \ref{figSV40} we portray the portion of the assembly tree  for SV40 limited to
species with up to 16 building blocks.  
%----------------------------------------- Figure --------------------
%---------------------------------------------------------------------
It contains 19 assembly intermediates, out of a {\it total number} of
505 assembly intermediates that are encoded by the minimal free-energy
assembly tree \cite{KTT}. The complicated structure of the assembly
tree makes impractical the use of relation (\ref{ieq}) for computing
the relative concentrations of the assembly intermediates. Therefore,
only the concentrations of the dominant assembly intermediates,
i.e. those located on all paths in the assembly tree such as, for
example, the intermediates denoted as {\bf 1},\,01 to {\bf 5},\,01 and {\bf 8},\,01 to {\bf 12},\,01 in
Fig. \ref{figSV40}, have been computed prior to this work (see
ref. \cite{KTT}). However, the concentrations of all assembly
intermediates are needed in order to obtain clues about the putative
pathways of assembly, and hence about possible mechanisms for
anti-viral drug design.

To compute the equilibrium probabilities of occupation of the various
species, and to identify the dominant kinetic pathway within the
assembly tree of Fig. \ref{figSV40} we must go beyond the mere
specification of the adimensional parameters $a/c$ and $b/c$ and
consider the absolute values of the energies $a,\ b,\ c$. The absolute
value of the nominal free energies provided by the VIPER website are
more than an order of magnitude larger than the typical interaction
energies of biomolecules (usually of the order of a few $Kcal/mol$).
The particularly high VIPER values may indeed reflect the shortcoming
of the rigid-unit approximations involved in the potential extraction
scheme. We shall therefore obtain an indication of the absolute scale
for the SV40 free energies and of the concentration $c_0$, by
following some guidelines inspired by previous theoretical and
experimental work.

 The first quantitative experimental input pertains to the overall
concentration of fundamental units present in solution, $[c^*]$, which
has, typically, of the order of 10 $\mu$M. Secondly, as anticipated,
we wish to describe the situation where the dominant species in
equilibrium are $[1]$ and $[N]$ \footnote{For pentamers in solution we
do not distinguish between the two different types of building blocks
in Fig. \ref{BuildingBlocks} as their C-terminal arms are dangling
freely and the building blocks are a way of encoding local bonding
structures when bound in the capsid.}.  Assuming that $[N]_{eq}$ and
$[1]_{eq}$ are equiprobable one has:
\begin{equation}\label{eq:20}
1= \frac{[N]_{eq}}{[1]_{eq}} = 12\, ( {{\widetilde{[1]}_{eq}} \over 5})^{71} e^{\kappa \, a/RT}\,,
\end{equation} 
\noindent where $\kappa:= 180 + 60 b/a + 30 c/a \approx
360.8571$\footnote{Note that this equation relates the association
energy $a$ with the equilibrium concentrations of pentamers,
$[1]_{eq}$, and hence changes in $[1]_{eq}$ may be engineered by
changing $a$. The latter can be achieved for example via alterations
in the polypeptide chain of the proteins (see
e.g. \cite{Zlotnick:Nano}).}.

The above requirement provides a relationship through the free energy
scale, $a$, and $c_0$ (entering implicitly through
$\widetilde{[1]}_{eq})$. The second condition typing $c_0$ and $a$ is
obtained by requiring that the concentration of species $[2]$ is
significantly smaller than $[1]$. This requirement implies, through
the chain relations of equation \ref{ieq}, that the dominant species are indeed
$[1]$ and $[N]$, so that

\begin{equation}
[c^*] = \sum_{k=1}^{505} n_k \frac{5^{n_k}}{O(k)} e^{- \Delta G(1,k) \over R T} 
{\widetilde{[1]}_{eq}}^{n_k-1}[1]_{eq} \approx [1]_{eq} + 72\, [N]_{eq} = 73\, [1]_{eq}\ .
\end{equation}

We discuss here the case where $[2] = [1]/10$, which is satisfied when
$a$ takes on the realistic value, $a \approx -0.7 kcal/mol$. All our
conclusions about the dominant pathway in the assembly tree are
unchanged if much larger values of $a$ (in modulus) are used, although
these might result in unrealistically low concentrations of
intermediates.

Therefore the assembly tree for SV40 has been computed for the values
$a=-0.7$\,kcal\,mol$^{-1}$, $b=-1.37$\,kcal\,mol$^{-1}$ and
$c=-1.49$\,kcal\,mol$^{-1}$.  The shortest pathways in the tree that
connect $[1]$ and $[N]$ (i.e. those without loops) contains precisely
72 species, one for each possible value of building blocks, see
Fig. \ref{BuildingBlocks}.

\begin{figure}[ht]
\begin{center}
\includegraphics[width=10cm,keepaspectratio]{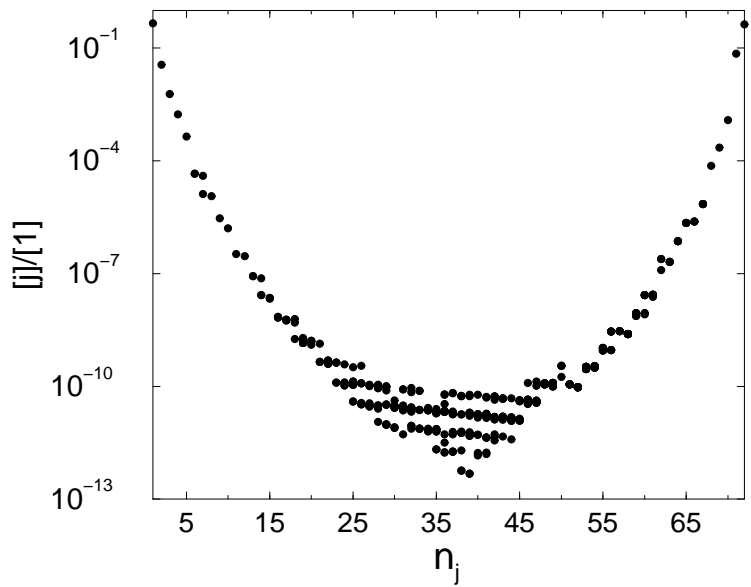}
\end{center}
\caption{SV40 capsid assembly: scatter plot for the concentrations of
  assembly intermediates, $[j]$, as a function of the number of
  building blocks, $n_j$. The concentrations were normalised with
  respect to that of the fundamental unit, $[1]$. Notice that more
  than a species may exist for given $n_j$. The plot refers to the
  situation where $[c^*] = 10 \mu$M and $a=-0.7 kcal/mol$.}
\label{ProbDist}
\end{figure}

We have computed the concentrations of the assembly intermediates in
thermodynamic equilibrium as shown in Fig. \ref{ProbDist}.
%----------------------------------------- Figure --------------------
%--------------------------------------------------------------------

In particular, one observes that concentrations are highest at the
beginning and at the end of the assembly pathways, and are strictly
and rapidly decreasing (at the beginning) or increasing (at the
end). For the intermediates at the start of the assembly tree shown in
Fig. \ref{figSV40} one observes furthermore the following scenario: in
the cases where more than one intermediate of the same number of
building blocks exists (such as for example {\bf 6},\,01 and {\bf 6},\,02, or, {\bf 7},\,01 and
{\bf 7},\,02) their concentrations are either identical as a consequence of
degeneracy (such as for {\bf 6},\,01 and {\bf 6},\,02 and all other nodes springing out
from a parent node), or vary strongly ({\bf 7},\,02 having a larger
concentration than {\bf 7},\,01). In the latter case, we indicate the
intermediate with the larger concentration by a double circle in
Fig. \ref{figSV40}. Our computations show furthermore that for
comparable values of probabilities towards the end of the pathways,
different intermediates with the same number of building blocks (and
hence all assembly pathways) are indistinguishable. Therefore, the
start of the assembly pathways singles out the dominant pathway(s).
Since dead-ends and traps do not occur in the assembly tree, the
pathways containing the two largely dominant configurations {\bf 7},\,02 and
{\bf 14},\,02 in Fig. \ref{figSV40} must be the dominant pathways during
assembly.

\begin{figure}[ht]
\begin{center}
\includegraphics[width=4cm,keepaspectratio]{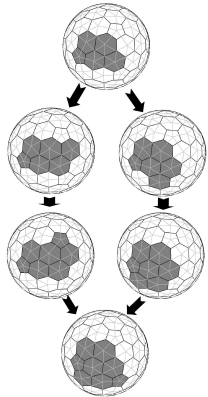}
\end{center}
\caption{The geometries of the intermediates {\bf 7},\,02 and {\bf 14},\,02.}
\label{geom}
\end{figure}

The geometries of the intermediates {\bf 7},\,02 and {\bf 14},\,02 are shown in
Fig. \ref{geom}. 
%----------------------------------------- Figure --------------------
%--------------------------------------------------------------------
One observes that in each case, the dominant configuration
is obtained from the previous intermediate via the formation of bonds
with association energy $2a$, $a$ and $b$. The fact that the formation
of this constellation of bonds is important is corroborated further by
the following: We have increased the association energies of the bonds
$a$ and $b$ individually, and have compared the ratio of final capsids
to pentamers in equilibrium. In both cases the yield of final capsids
has increased, with a stronger increase in response to an increase of
the association energy related to the bond with association energy
$b$.
\medskip

These considerations suggest that SV40 capsid assembly is driven more
by the details of the association free energies rather than
differences in the geometrical entropy associated to rotational
symmetries of the various species. This fact can be conveniently
checked by setting to zero all association free energies, $a,\ b$ and
$c$ and computing the contribution of the factors $S_1$ and $S_n$ to
the concentrations of the various species. One observes that all
assembly intermediates (on different assembly pathways) of an equal
number of building blocks have the same probability, with the
exception of assembly intermediates that have a discrete rotational
symmetry. However, since these symmetry corrections appear only at
later stages in the pathways (first occurrence at iteration step 30)
where concentrations are low when energy effects are counted in, they
do not need to be considered when distinguishing the dominant
pathways. The dominant pathways are, therefore, strongly dependent on
the association free energies.
\bigskip

\begin{figure}[h]
\begin{center}
\includegraphics[width=3cm,keepaspectratio]{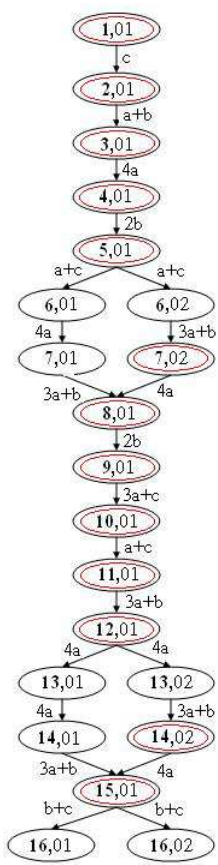}
\end{center}
\caption{The start of the assembly tree for a representative of a
different area in phase space.}
\label{figBOX1}
\end{figure}

In \cite{KTT} it has been shown that the phase space formed by the
ratios of the association constants can be partitioned into areas in
which the qualitative behaviour of assembly, as encoded in the
assembly tree, is indistinguishable. In order to explore if similar
results occur also for these other areas in phase space (and hence for
different assembly trees), we have calculated the concentrations of
the assembly intermediates for a representative of a different area in
the partition (area 1 in Fig.~5 \cite{KTT}). It corresponds to the
point $x=0.75$ and $y=0.45$ in phase space. The complete assembly tree
consists of 281 species, and the start of the assembly tree is shown
in Fig. \ref{figBOX1}.
%----------------------------------------- Figure --------------------
%---------------------------------------------------------------------  
The assembly intermediates with the larger concentrations are again
marked by double circles. As in the case of SV40 assembly, the
occurrence of assembly intermediates with concentrations larger that
that of the other intermediates with the same number of building
blocks is related to the formation of bonds with association energies
$2a$, $a$ and $b$. This phenomenon hence seems pertinent to the
selection of the dominant pathways, and may therefore provide insight
into the aspect of the viral assembly which may become target of
drugs.

\section{Discussion}

We have demonstrated that a combination of the master equation and the
tiling approach allows to determine the putative pathways for SV40
capsid assembly and sheds light on the mechanisms that drive capsid
assembly. In particular, we have demonstrated that the more important
assembly pathways are those where a particular constellation of bonds
is formed at an early stage (see Fig. \ref{geom}). Hence, this constellation
of bonds could be a possible target for antiviral drug design; for
example, it suggests to search for a drug that binds to the sites
related to these bonds, hence preventing the formation of this
particular constellation of bonds.

Moreover, our analysis has shown that SV40 capsid assembly is strongly
driven by the details of the association free energies of the tiles
and is only slighted affected by the entropic aspects associated to
the different rotational symmetries of the various species.  While
this may be similar for other DNA viruses, it is presumably not the
case for RNA viruses due to the interactions between RNA and the
protein building blocks of the capsids. This fact justifies to neglect
other combinatorially possible intermediates, which however may be
important for other viruses as demonstrated in \cite{Endres:2005}.

\section*{Acknowledgements}
We are indebted to Adam Zlotnick for valuable comments in response to
a careful reading of the manuscript.  RT has been supported by an
EPSRC Advanced Research Fellowship. TK has been supported by the EPSRC
grant GR/T26979/01.


\begin{thebibliography}{99}

\bibitem{Berger:1994}
Berger, B. et al. (1994), {\it Proc. Natl. Acad. Sci.} {\bf 91}, pp. 7732.

\bibitem{Brooks:1998}
Brooks III, C.L. et al. (1998), {\it Proc. Natl. Acad. Sci.} {\bf 95}, pp. 11037.

\bibitem{Casjens:1985}
Casjens, S. (1985). Virus structure and assembly. Jones and Bartlett, Boston, Massachusets.

\bibitem{Zlotnick:Nano}
Johnson, J.M. et al. (2005) - Nanoletters - complete... 

\bibitem{Klug:1962}
Caspar, D.L.D. and A. Klug (1962), {\it Cold Spring Harbor Symp. Quant. Biol.} {\bf 27}, pp. 1.

\bibitem{Endres:2002}
Endres, D. and A. Zlotnick (2002), {\it Biophysical Journal} {\bf 83}, pp. 1217.

\bibitem{Endres:2005}
Endres, D., Miyahara, M., Moisant, P.  and A. Zlotnick (2005), {\it Protein Science} {\bf 14}, pp. 1518.

\bibitem{Horton:1992}
Horton, N. and M. Lewis (1992), {\it Protein Science} {\bf 1}, pp. 169.

\bibitem{Itzykson}
Itzykson \& Drouffe, Statistical field theory, 2nd volume

\bibitem{KTT} Keef, T., Taormina, A. and Twarock, R. (2005) {\it
Assembly Models for Papovaviridae based on Tiling Theory}, submitted
to {\it J. Phys. Biol.}.

\bibitem{Liddington:1991}
Liddington, R.C. et al. (1991), {\it Nature}  {\bf 354}, pp. 278.

\bibitem{Modis:2002} 
Modis, Y., et al. (2002), {\it EMBO J.} {\bf 21}, pp. 4754. 

\bibitem{Rapaport:1999} 
Rapaport, D.C., et al. (1999), {\it Comp. Phys. Comm.} {\bf 121}, pp. 231. 

\bibitem{Rapaport:2004} 
Rapaport, D.C., et al. (2004), {\it Phys. Rev. E} {\bf 70}, pp. 

\bibitem{Rayment:1982}
Rayment, I., et al. (1982), {\it Nature} {\bf 295}, pp. 110. 

\bibitem{Reddy:1998}
Reddy, V. S. et al. (1998), {\it Biophysical Journal} {\bf 74}, pp. 546. 

\bibitem{Reddy:2001}
Reddy, V. S. et al. (2001). {\it J. Virol.} {\bf 75} pp. 11943. 

\bibitem{Schwartz:2000}
Schwartz, R. et al. (1998). {\it Biophys. J.} {\bf 75} pp. 2626. \\
Schwartz, R. et al. (2000). {\it Virology} {\bf 268} pp. 461. 

\bibitem{Twarock:2004a}
Twarock, R.  (2004), {\it J. Theor. Biol.} {\bf 226}, pp. 477. 

\bibitem{Twarock:2004b}
Twarock, R. (2005), {\it Bull. Math. Biol.} {\bf 68}, in press. 

\bibitem{Wales:1987}
Wales, D.J. (1987), {\it Chem. Phys. Lett.} {\bf 141}, pp. 478. 

\bibitem{Wales:1996}
Wales, D.J. (1996), {\it Science} {\bf 271}, pp. 925. 

\bibitem{Wolynes:1996}
Wolynes, P.G. (1996), {\it Proc. Natl. Acad. Sci.} {\bf 93}, pp. 14249.

\bibitem{Zandi:2004}
Zandi, R. et al. (2004), {\it Proc. Natl. Acad. Sci.} {\bf 101}, pp. 15556. 

\bibitem{Zlotnick:1994}
Zlotnick, A. (1994), {\it J. Mol. Biol.}, {\bf 241}, pp. 59. 

\bibitem{Zlotnick:1999}
Zlotnick, A. et al., (1999), {\it Biochemistry}, {\bf 38}, pp. 14644. 

\bibitem{Zlotnick:2000}
Zlotnick, A. et al., (2000), {\it Virology}, {\bf 277}, pp. 450. 

\bibitem{Zlotnick:2002}
Zlotnick, A. et al., (2002), {\it J. Virol.}, {\bf 76}, pp. 4848. 

\end{thebibliography}
\end{document}